\documentclass[aps,prb,twocolumn,superscriptaddress]{revtex4-2}
\usepackage{amsmath,amsthm,amssymb,bm}
\usepackage{graphicx}
\usepackage{chemformula}
\usepackage{xcolor}

\begin{document}

\newcommand{\bi}[1]{\ensuremath{\boldsymbol{#1}}}

\title{Electron $g$-value comparison of paramagnetic muonium and hydrogen centers\\ in rutile TiO$_2$}

\author{T.~U.~Ito}\email[tuito@post.j-parc.jp]\
\affiliation{Advanced Science Research Center, Japan Atomic Energy
 Agency, Tokai, Ibaraki 319-1195, Japan}
\author{W.~Higemoto}
\affiliation{Advanced Science Research Center, Japan Atomic Energy
Agency, Tokai, Ibaraki 319-1195, Japan}
\affiliation{Department of Physics, Institute of Science Tokyo,
Meguro, Tokyo 152-8551, Japan}
\author{A.~Koda} 
\affiliation{Institute of Materials Structure Science, High Energy
Accelerator Research Organization (KEK), Tsukuba,
Ibaraki 305-0801, Japan}

\date{\today}

\begin{abstract}
Muonium (Mu) is widely used as a model for isolated hydrogen in
condensed matter, but whether its electronic state is truly equivalent
  to that of hydrogen remains to be tested with an orbital-sensitive probe.
Here we use double electron-muon resonance to determine the electron $g$ 
values of the paramagnetic Mu center in rutile TiO$_2$ for magnetic
 fields along [100], [110], and [001] directions.
The obtained $g$ values closely follow those of the H-related Ti$^{3+}$ 
paramagnetic center in both magnitude and anisotropy.
This agreement tests the Mu--H correspondence at the level of the Ti 
$3d$ electronic state and supports a common localized Ti$^{3+}$ 
electronic core for the two centers.

 \end{abstract}

\maketitle

Muonium ($\mathrm{Mu}$) has long been used as a model for hydrogen
impurities in semiconductors and wide-gap oxides to investigate their
various electronic states, local bonding configurations, and dynamics
via muon spin rotation, relaxation, and resonance ($\mu^+\mathrm{SR}$)
spectroscopy~\cite{patterson88,cox06_1,cox06_2,cox09,ito20}.
Particularly for characterizing paramagnetic defects (i.e., neutral
$\mathrm{Mu}^0$ states), most electronic and structural assignments have
relied on the $\mathrm{Mu}$ hyperfine parameters, whose magnitude,
anisotropy, and temperature dependence are compared with models or, when
available, with magnetic resonance data for corresponding H$^0$ defects.
The hyperfine interaction is a powerful local probe, but it is not a
purely electronic fingerprint: it is also sensitive to the muon/proton
density distributions including the effects of zero-point motion, as
well as their local dynamics~\cite{vilao15,dehn20}.
A stringent test of the electronic part of the
$\mathrm{Mu}$--$\mathrm{H}$ analogy in these paramagnetic systems
therefore requires a complementary observable tied more directly to the
orbital state of the unpaired electron.
The electronic $g$ value can fulfill this role, as it is highly
sensitive to the local crystal-field splitting of electron orbitals and
spin--orbit coupling~\cite{yang13}.

Rutile $\mathrm{TiO}_2$ serves as an ideal platform for such a precise
comparison because the hyperfine parameters and $g$ values of a
paramagnetic H$^0$ defect have been thoroughly characterized by electron
paramagnetic resonance (EPR) and electron-nuclear double resonance
(ENDOR)~\cite{brant11}.
For both tensors, two of the principal axes lie within the $ab$ plane,
while the third runs parallel to the $c$ axis.
Since the three principal values of the hyperfine tensor ($A_1, A_2,
A_3$) roughly follow a $-1:+2:-1$ ratio, the magnetic interaction is
predominantly dipolar.
This indicates that the unpaired electron is primarily localized 
at a single Ti site separated from the proton by an atomic-scale
distance within the same $ab$ plane, thereby reducing a host
$\mathrm{Ti}^{4+}$ ion to a nominally $\mathrm{Ti}^{3+}$ state. 
The angular dependence of the $g$ value is likewise
consistent with a model incorporating crystal-field splitting and
spin--orbit coupling for an excess electron localized in the Ti $3d$
orbital.

For the $\mathrm{Mu}^0$ defect, however, independent studies based on
hyperfine parameters have yielded qualitatively different interpretations.
Vil{\~a}o~\textit{et al.} showed that the $\mathrm{Mu}^0$ hyperfine
splitting under a transverse magnetic field (TF) along the $[110]$
direction, when scaled by the respective gyromagnetic ratios, closely
matches that of the $\mathrm{H}^0$ defect~\cite{vilao15}.
On this basis, they argued that $\mathrm{Mu}^0$ and $\mathrm{H}^0$ form
the same configuration with the same basic electronic structure,
supporting the picture of a $\mathrm{Ti}^{3+}$ small polaron bound to a
$\mathrm{Mu}^+$ donor.
In contrast, when the TF is applied along $[001]$, no clear line
splitting was observed, meaning the corresponding principal value $A_3$ is nearly
zero. Thus, from the hyperfine perspective, the
$\mathrm{Mu}$--$\mathrm{H}$ correspondence remains incomplete, and the
$A_3$ discrepancy was ascribed to the larger zero-point motion of 
the muon along the $c$ direction than that of the proton.
While these comparisons apply to the ground-state $\mathrm{Mu}^0$ at
1.2~K, warming the system above $\sim$4~K induces a partial merging of
the hyperfine splitting, which has been attributed to thermal excitation
of a metastable Mu$^0$ state and subsequent dynamic averaging of the hyperfine
parameters.
On the other hand, Shimomura~\textit{et al.} investigated the angular
dependence of the hyperfine splitting at 5~K~\cite{shimomura15}.
Analyzing the signal under the assumption of a single static $\mathrm{Mu}^0$
configuration, they proposed a large-polaron-like picture, in which 
 the effective spin density is substantially reduced at the nearest Ti
 site and distributed over a broader Ti--O environment.

In this Letter, we report on double electron-muon resonance (DEMUR) 
measurements of the $g$ values for the paramagnetic $\mathrm{Mu}^0$
defect in rutile $\mathrm{TiO}_2$.
DEMUR measurements using a radio-frequency (rf) $\mu^+$SR probe provide
an orbitally sensitive metric to
determine whether $\mathrm{Mu}^0$ and $\mathrm{H}^0$ share the same
underlying electronic state~\cite{brown83,estle83,blazey86,lord04,doll25}.
From measurements at 2.5~K, we determined the $g$ values for the
ground-state $\mathrm{Mu}^0$ along the $[100]$, $[110]$, and $[001]$
field directions, and found that they closely follow those of the
corresponding $\mathrm{H}^0$ center.
This agreement tests the $\mathrm{Mu}$--$\mathrm{H}$ correspondence at
the level of the $\mathrm{Ti}$ $3d$ electronic state and supports a
common localized $\mathrm{Ti}^{3+}$ electronic core for the
$\mathrm{Mu}^0$/$\mathrm{H}^0$ centers.

Throughout this Letter, Mu$^0$, H$^0$, and Li$^0$ denote charge-neutral
paramagnetic defect complexes. In Kr\"{o}ger--Vink
notation~\cite{kroger56,norby10}, these centers
may be represented schematically as
$\left(X_i^{\bullet}{\rm Ti}_{\rm Ti}^{\prime}\right)^{\times}$, i.e., an
ionized interstitial donor component $X$ compensated by a Ti$^{3+}$ small
polaron. The labels Mu$^+$, H$^+$, and Li$^+$ are used only as formal
notations for the positively charged donor components in this
defect-chemical picture. In particular, for H and Mu these donor
components are structurally realized as oxygen-bound OH and OMu
groups~\cite{brant11,vilao15,shimomura15}.  By contrast, Ti$^{3+}$ and
Ti$^{4+}$ denote formal Ti oxidation states.

We adopted a TF implementation of DEMUR in the decoupling regime reached
at high rf power, 
following the method introduced by Lord {\it et al.} for investigating 
shallow-donor muonium centers~\cite{lord04}. In this implementation, the
rf field is applied at
the EPR frequency of the Mu$^0$ center in rutile TiO$_2$
under a static TF $\bi{B_0}$,
while the resonance is detected through changes in the time-differential
TF-$\mu^+$SR signal. The relevant regime is reached when the amplitude of
the rf field $B_1$ satisfies $g\mu_{\rm B}B_1/h \gtrsim A$, where $g$ is
the directional $g$ value of the unpaired electron, $\mu_{\rm B}$ is the
Bohr magneton, $h$ is Planck's constant, and $A$ ($\sim$1.5~MHz for
Mu$^0$ in rutile TiO$_2$) denotes the
characteristic electron--muon hyperfine coupling scale. Under this condition, 
resonant driving of the electron spin occurs on a time scale shorter
than that set by the hyperfine interaction. The electron resonance is therefore
detected through the rf-induced dynamical averaging of the hyperfine
response in the TF-$\mu^+$SR lineshape, rather than through perturbative
shifts, splittings, or intensity redistributions of individual muon
precession components, as in earlier DEMUR studies outside this
decoupling regime~\cite{brown83,estle83,blazey86}.

The TF-DEMUR measurements in the decoupling regime on
single-crystalline rutile TiO$_2$ were performed at the Materials and Life Science
Experimental Facility (MLF) of the Japan Proton Accelerator Research Complex
(J-PARC) using the S1 ARTEMIS $\mu^+$SR spectrometer equipped with
1280-channel solid-state positron detectors~\cite{kojima14,strasser18}.
Nominally undoped rutile TiO$_2$ substrates grown by the Verneuil method
were obtained from Furuuchi Chemical Co., Japan.  Each substrate had
dimensions of $10\times10\times0.5$~mm$^3$.  To increase the effective
sample area exposed to the muon beam, nine substrates were tiled in a
$3\times3$ array.  For measurements with $\bi{B_0}\parallel[100]$
and $[001]$, we used $b$-plane substrates, with
their in-plane crystallographic axes carefully aligned across the array.
The static TF was then applied parallel to either the $a$ axis
$\parallel[100]$ or the $c$ axis $\parallel [001]$.  For measurements with
$\bi{B_0}\parallel[110]$, we used an analogous $3\times3$ array of
$c$-plane substrates with the in-plane $[110]$ direction aligned
parallel to $\bi{B_0}$.
The $3\times3$ substrate array was mounted inside a ten-turn flat rf coil
with dimensions of $40\times30\times3$~mm$^3$.  The coil was installed
at the end of the sample stick of a $^4$He-flow cryostat and 
incorporated into an rf tank circuit equipped with variable capacitors
for frequency tuning and impedance matching.
The static TF $\bi{B_0}$ and the rf field $\bi{B_1}$ generated by the
coil both lay in the sample plane and were mutually perpendicular, while
the initial muon spin polarization was normal to
the sample plane.

Spin-polarized pulsed surface muons were delivered to the sample at a
repetition rate of 25~Hz. The rf condition was alternated between on and
off for successive muon pulses.  For pulses assigned to the rf-on
condition, a 60.000-MHz rf field with an amplitude $B_1$ 
$\sim0.1$~mT was applied for 25~$\mu$s, starting 2.7~$\mu$s before
the arrival of the corresponding muon pulse.  Muon-decay events recorded
by the forward and backward positron detectors were sorted into separate
histograms according to the rf condition.  The rf-on and rf-off
histograms were converted separately into the asymmetry, $A(t)=[N_{\rm
F}(t)-\alpha N_{\rm B}(t)]/[N_{\rm F}(t)+\alpha N_{\rm B}(t)]$, 
where $N_{\rm F}(t)$ and $N_{\rm B}(t)$ are the positron counts in the forward and
 backward detectors as functions of the time $t$ elapsed from the
 instant of muon implantation, and $\alpha$ is a parameter to correct for
 the counting efficiencies of the forward and backward detectors. In
 particular, $A(t)$ is referred to as the corrected asymmetry when
 $\alpha$ is adjusted to balance the efficiencies at 1:1, whereas it is
 termed the raw asymmetry when $\alpha=1$.

 The DEMUR response was then evaluated by directly comparing the rf-on and
rf-off asymmetries accumulated within the same measurement run.
For each field direction, $B_0(=|\bi{B_0}|)$ was swept at
the fixed rf frequency of $\nu_{\rm rf}=60.000$~MHz with the sample
 held at 2.5~K.  Within each field sweep, the counting time at
each field point was adjusted so that the total number of positron
events remained approximately constant.
 The resonance field $B_0^{\rm res}$ was determined from the field
 dependence of the rf-induced change in the
 TF-$\mu^+$SR lineshape.  The directional $g$ value was obtained
from the resonance condition

\begin{equation} 
 \label{eq1}
 h\nu_{\rm rf}=g\mu_{\rm B}B_0^{\rm res}.
\end{equation}
The actual value of $B_0$ at each nominal field point was
independently calibrated from the muon spin precession frequency measured
in conventional TF-$\mu^+$SR spectra at 100~K, where only the diamagnetic
muon signal was observed.

\begin{figure*}[tb]
\includegraphics[scale =0.6]{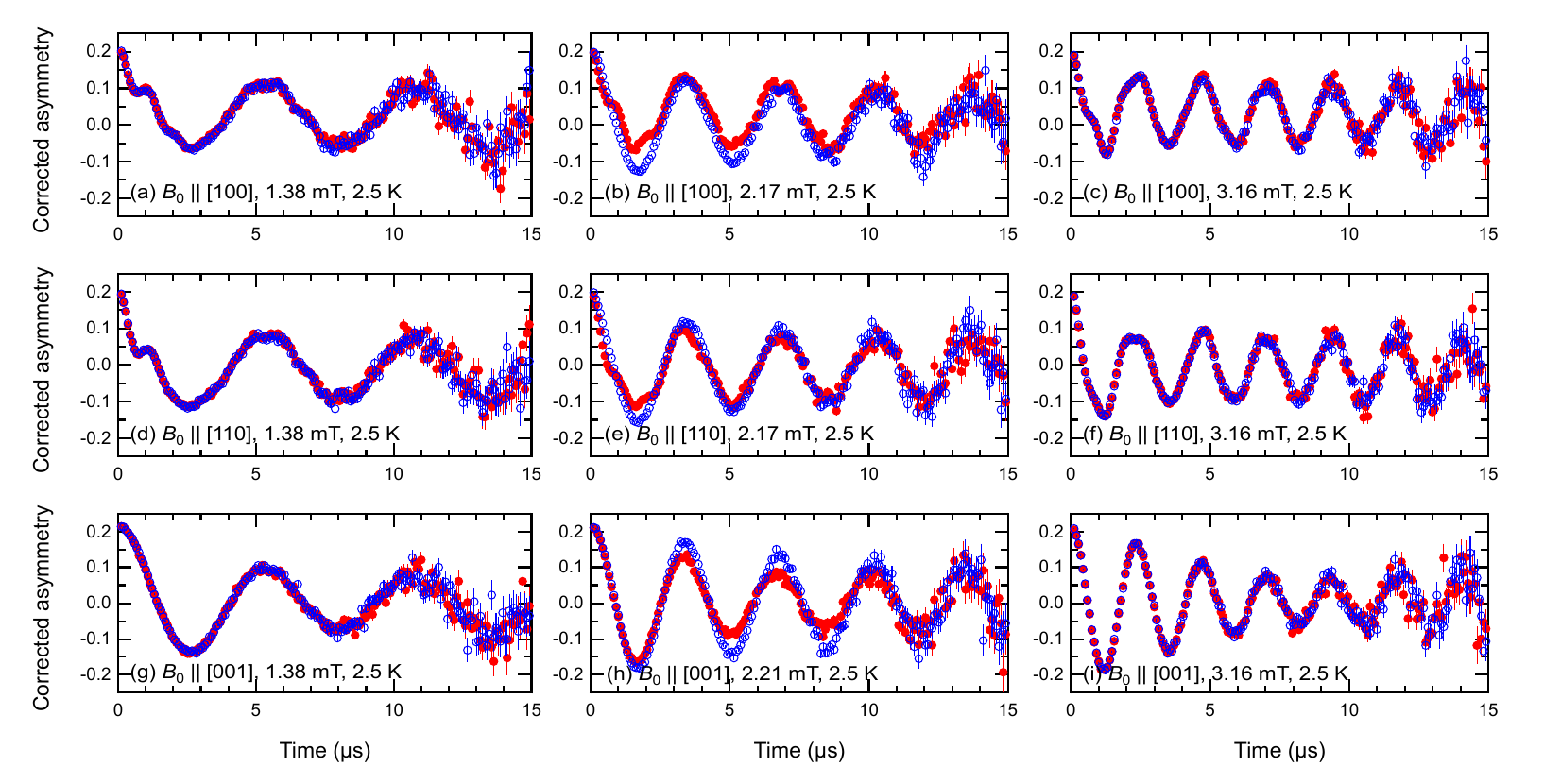}
\caption{\label{fig1}Comparison of the corrected asymmetries at 2.5~K
 under rf-on (blue open circles) and rf-off (red closed
  circles) conditions. Panels (a--c), (d--f), and (g--i) correspond to
 $\bi{B_0} \parallel [100]$, $\bi{B_0} \parallel [110]$, and $\bi{B_0}
 \parallel [001]$, respectively. From left to right, $B_0$
 increases from 1.38~mT to 3.16~mT, with the middle panels (b, e, h) taken near 
 $B_0^{\mathrm{res}}$ for each field direction.}
\end{figure*}

Figure~\ref{fig1} compares the time-differential corrected asymmetries
measured under the rf-on and rf-off conditions.  For all three
orientations of $\bi{B_0}$, a clear difference between the rf-on and
rf-off data appears around $B_0=2.2$~mT, demonstrating that the
on-resonance condition [Eq.~\ref{eq1}] is fulfilled.  Under the rf-on condition,
the fine oscillatory structure arising from the hyperfine interaction is
strongly suppressed, leaving a slowly damped cosine curve.
This behavior is consistent with dynamical averaging of the hyperfine
interaction in the decoupling regime.  By contrast, at
fields well below and above resonance, such as $B_0=1.38$~mT and
$3.16$~mT, respectively, no discernible difference is observed between
the rf-on and rf-off lineshapes.

To quantify the resonance response from this series of TF-$\mu^+$SR
spectra, it is useful to reduce the rf-induced change in $A(t)$ to a
scalar quantity.  A straightforward choice would be time-integrated
rf asymmetry, as commonly used in conventional muon spin resonance
measurements in a longitudinal field geometry~\cite{kreitzman95,scheuermann97}.
However, as shown in the Supplemental 
Material, this quantity does not provide a reliable measure of the
present TF-DEMUR response~\cite{sm}.  Another possible approach, following the
analysis used by Lord {\it et al.}, would be to extract the apparent
hyperfine splitting of Mu$^0$ signals by curve fitting and use its
rf-induced reduction as the resonance indicator~\cite{lord04}.  In the present
experiment, however, this procedure is not well suited because the
measurements are performed in a low field region where the Mu$^0$
frequencies vary nonlinearly with $B_0$.

\begin{figure}
 \includegraphics[scale =0.6]{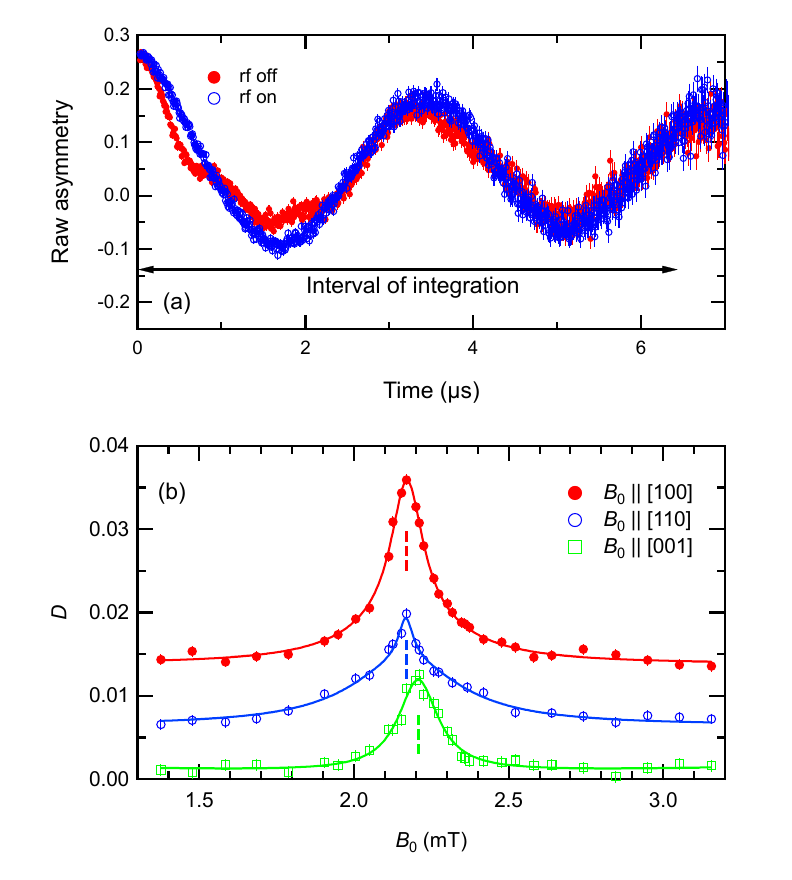}
   \caption{\label{fig2} (a) Raw asymmetries with 8~ns per bin for $\bi{B_0} \parallel [110]$ at $B_0 = 2.17$~mT, and
 $T = 2.5$~K, illustrating the interval of integration in the definition
 of $D$ [Eq.~(\ref{eq2})].
 (b) $B_0$ dependence of $D$ for $\bi{B_0} \parallel [100]$, $[110]$, and $[001]$.
 The data for $\bi{B_0} \parallel [110]$ and $[001]$ are
 vertically offset by $-0.007$ and $-0.014$, respectively. Solid
 curves represent the best fits using a double-Lorentzian function with
 a shared center and a constant background term. Vertical broken lines indicate the resonance fields
 corresponding to the $g$ values of the H$^0$ center~\cite{brant11}.}
 \end{figure}

We therefore adopted a fitting-free measure of the rf-induced 
change in $A(t)$, defined as the time-averaged absolute difference, 
\begin{equation}
 \label{eq2} 
 D=\frac{1}{t_{\rm cut}}\int_{0}^{t_{\rm cut}}\left| A_{\rm raw}^{\rm on}(t)-A_{\rm raw}^{\rm off}(t) \right|dt,
\end{equation}
where $A^{\rm on}_{\rm raw}(t)$ and
$A^{\rm off}_{\rm raw}(t)$ denote the time-differential raw
asymmetries with the minimum time-bin width of 8~ns under the
rf-on and rf-off conditions, respectively.
For the uniform time-bin width used here, $D$ is equivalent to the
arithmetic mean of the absolute asymmetry differences over the time bins
within the averaging interval.
The upper limit of the averaging interval
 was set to $t_{\rm cut}=6.4~\mu$s, based on the typical time at which
the absolute difference becomes sufficiently small for
$\bi{B_0}\parallel[110]$, as shown in Fig.~\ref{fig2}(a).
By evaluating this quantity, we obtained the $B_0$ dependence of
$D$ for $\bi{B_0}\parallel[100]$, $[110]$, and $[001]$, as shown in
Fig.~\ref{fig2}(b).
The solid curves represent the best fits to the datasets
using a double-Lorentzian function with a shared center ($B_0^{\rm
res}$) and a constant background term.
The center of each resonance line for Mu$^0$ was 
found to agree well with the corresponding $B_0^{\rm res}$ value for
H$^0$ calculated from its directional $g$ value~\cite{brant11}.
Using the fitted values of $B_0^{\rm res}$ and
Eq.~(\ref{eq1}), we obtained $g_{[100]}=1.9741(10)$, $g_{[110]}=1.9776(25)$, and
$g_{[001]}=1.9417(25)$ for the ground-state Mu$^0$ center.

\begin{figure}
\includegraphics[scale =0.6]{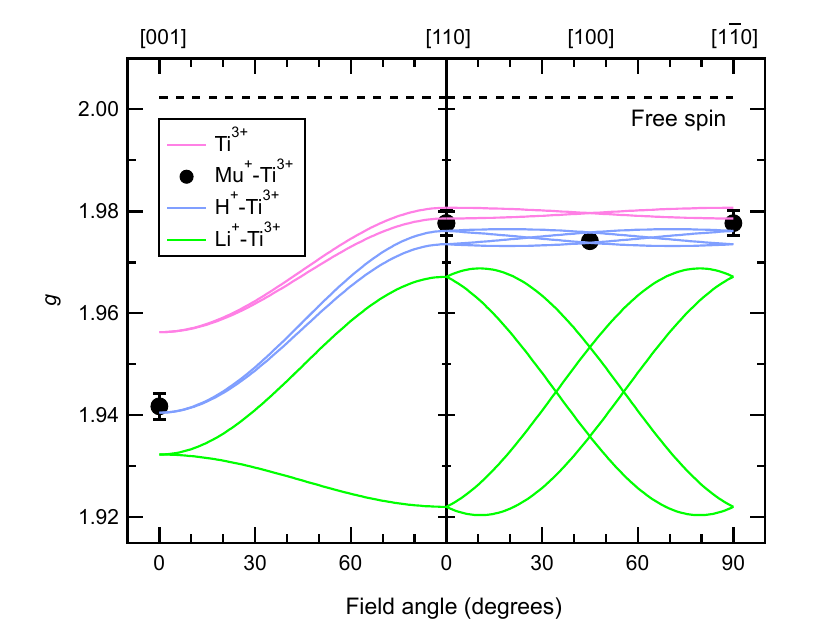}
\caption{\label{fig3} Field-angle dependence of the $g$ value for the
ground-state $\mathrm{Mu}^0$ center ($\mathrm{Mu}^+$--$\mathrm{Ti}^{3+}$
complex), along with those of other Ti$^{3+}$-related paramagnetic
centers, including a self-trapped Ti$^{3+}$ center~\cite{yang13} and Ti$^{3+}$
complexes associated with interstitial H$^+$ and Li$^+$
donors~\cite{brant11,brant13}.}
\end{figure}

Figure~\ref{fig3} provides a visual comparison of the magnitude and
anisotropy of the $g$ values for Mu$^0$ with those of Ti$^{3+}$-related
paramagnetic centers, including a self-trapped Ti$^{3+}$
center~\cite{yang13} and Ti$^{3+}$ complexes associated with
interstitial H$^+$ and Li$^+$ donors~\cite{brant11,brant13}.
A numerical comparison of these $g$ values along the
[100], [110], and [001] field directions is also given in the
Supplemental Material~\cite{sm}.
The good agreement in both magnitude and anisotropy between the $g$
values of the Mu and H-related centers provides a constraint that is
complementary to the comparison based on hyperfine parameters.  The $g$
tensor of a Ti$^{3+}$ small polaron is mainly
determined by the crystal field and spin--orbit coupling acting on the
localized Ti $3d$ electron. The present result therefore shows that
the unpaired electron in the Mu$^0$ center occupies essentially the same Ti
$3d$ orbital state as in the H$^0$ center with the localized Ti$^{3+}$
electronic core (H$^+$--Ti$^{3+}$ complex). This is a more direct test
of the electronic part of the Mu--H analogy than the hyperfine
interaction alone, because the hyperfine parameters are also affected by
the probability density distribution and dynamics of the light {\it nuclei}.

This conclusion should be viewed in light of the problem emphasized by
Shimomura {\it et al.}~\cite{shimomura15}.
They pointed out that the hyperfine scales observed by $\mu^+$SR for
Mu$^0$ ($\sim$1.5~MHz) and by ENDOR for H$^0$ ($\sim$0.5~MHz) are about
one order of magnitude smaller
than expected from a classical dipolar estimate for the nearest-neighbor
H$^+$--Ti$^{3+}$ geometry proposed in Ref.~\cite{brant11}. 
This discrepancy was one of the main motivations for their 
large-polaron-like picture, in which the effective spin density is
substantially reduced at the nearest Ti site and distributed over a
broader Ti--O environment.
Notably, the $\mu^+$SR hyperfine scale for rutile TiO$_2$ is roughly ten
times smaller than that for SrTiO$_3$~\cite{salman14}, where the
Mu$^{+}$--Ti$^{3+}$ complex has been well established~\cite{ito19,ito22}.
A possible way to reconcile the localized electronic core inferred from
the $g$ tensors with the small hyperfine scales observed in rutile
TiO$_2$ is that the electron is trapped on a Ti site located farther
from the ionized Mu$^+$ or H$^+$ donor than assumed in the simplest
nearest-neighbor geometry.
This would reduce the dipolar hyperfine field at the
muon or proton without requiring a broad distribution of the Ti $3d$
electron over many Ti sites.

The comparison with the Li$^0$ center (Li$^+$--Ti$^{3+}$ complex) is
suggestive in this respect.
In contrast to the Mu$^0$ and H$^0$ centers, the reported $g$ values of
the Li$^0$ center~\cite{brant13} deviate more strongly from those of
the self-trapped Ti$^{3+}$ center~\cite{yang13}, as shown in
Fig.~\ref{fig3}. The substantially larger hyperfine scale of the
Li-related center ($\sim$3~MHz) points to a shorter donor--Ti$^{3+}$
separation, as expected for a nearest-neighbor geometry~\cite{brant13}.
These features suggest that a nearby ionized donor can perturb the
crystal field more strongly.
Conversely, the smaller hyperfine interactions and the less perturbed
$g$ tensors of the Mu$^0$ and H$^0$ centers are consistent with a more
weakly coupled geometry, in which the positive donor and the Ti$^{3+}$
electron are separated by a somewhat larger distance.

Muon-specific electronic structure calculations have become an important
complement to $\mu^+$SR experiments, enabling microscopic analysis of muon
sites, local lattice relaxation, and muon-induced electronic
states~\cite{blundell25}. Combined experimental and computational
studies have provided microscopic descriptions of charge-neutral
muon--polaron complexes in transition-metal oxides such as Cr$_2$O$_3$, 
Fe$_2$O$_3$, and SrTiO$_3$~\cite{dehn20,dehn21,ito22}. The directional $g$ values
reported here offer a stringent experimental benchmark for future
calculations of the local electronic structure of Mu/H-related centers
in rutile TiO$_2$.

In summary, the present DEMUR measurements separate the electronic aspect
of the Mu--H analogy from the isotope-dependent hyperfine response.
The directional $g$ values support a common localized Ti$^{3+}$
electronic core for the ground-state Mu$^0$ and H$^0$ centers.  The
remaining differences in the hyperfine
parameters, especially for the principal component along the $c$ axis,
may be more naturally attributed to the quantum delocalization and/or  
dynamics of the Mu$^+$ and H$^+$ donors, rather than to a different
electronic state.
Thus, Mu$^0$ in rutile TiO$_2$ is a valid analogue of H$^0$ at the
level of the Ti $3d$ electronic state, while the hyperfine tensor
retains isotope-specific information.

\begin{acknowledgments}
We thank H.~Sakai, S.~Nishimura, and J.~G.~Nakamura for technical
 assistance, and K.~Shimomura, R.~Kadono, and K.~Fukutani for helpful
 discussions. The DEMUR experiments were conducted under 
 MLF user programs (Proposal Nos. 2020B0244 and 2024B0368).
 This research was supported by Grants-in-Aid (Nos. 24H00477,
 23K11707, and 20K12484) from the Japan Society for the Promotion of
 Science.
\end{acknowledgments}

\section*{Author Contributions}
T.U.I. conceived the study, developed the
rf probe, and, along with W.H. and A.K., carried out the DEMUR 
experiments. T.U.I. also performed data analysis and drafted the
manuscript. All authors discussed the results and approved the final
version of the manuscript.


\begin{thebibliography}{99}
\bibitem{patterson88} B.~D.~Patterson,
	 Muonium states in semiconductors, 
         Rev.~Mod.~Phys. {\bf 60}, 69 (1988). 
\bibitem{cox06_1} S.~F.~J.~Cox, J.~S.~Lord, S.~P.~Cottrell, J.~M.~Gil,
	H.~V.~Alberto, A.~Keren, D.~Prabhakaran, R.~Scheuermann, and
	A.~Stoykov,
	Oxide muonics: I. Modelling the electrical activity of hydrogen
	in semiconducting oxides, 
	J.~Phys.:~Condens.~Matter {\bf 18},
	1061 (2006). 
\bibitem{cox06_2} S.~F.~J.~Cox, J.~L.~Gavartin, J.~S.~Lord,
	S.~P.~Cottrell, J.~M.~Gil, H.~V.~Alberto, J.~Piroto~Duarte,
	R.~C.~Vil\~{a}o, N.~Ayres de Campos, D.~J.~Keeble, E.~A.~Davis,
	M.~Charlton, and D.~P.~van~der~Werf,
	Oxide muonics: II. Modelling the electrical activity of hydrogen
	in wide-gap and high-permittivity dielectrics, 
	J. Phys.: Condens. Matter {\bf 18},
	1079 (2006). 
\bibitem{cox09} S.~F.~J.~Cox,
	Muonium as a model for interstitial hydrogen in the
	semiconducting and semimetallic elements, 
        Rep.~Prog.~Phys. {\bf 72}, 116501 (2009). 
\bibitem{ito20} T.~U.~Ito, W.~Higemoto, and K.~Shimomura,
	Negatively charged muonium and related centers in solids,  
        J.~Phys.~Soc.~Jpn. {\bf 89}, 051007 (2020). 


\bibitem{vilao15} R.~C.~Vil\~ao, R.~B.~L.~Vieira, H.~V.~Alberto,
	J.~M.~Gil, A.~Weidinger, R.~L.~Lichti, B.~B.~Baker,
	P.~W.~Mengyan, and J.~S.~Lord,
	Muonium donor in rutile TiO$_2$ and comparison with hydrogen,
	Phys.~Rev.~B {\bf 92}, 081202(R) (2015).

\bibitem{dehn20} 
        M.~H.~Dehn, J.~K.~Shenton, S.~Holenstein, Q.~N.~Meier,
        D.~J.~Arseneau, D.~L.~Cortie, B.~Hitti, A.~C.~Y.~Fang,
        W.~A.~MacFarlane, R.~M.~L.~McFadden, G.~D.~Morris, Z.~Salman,
        H.~Luetkens, N.~A.~Spaldin, M.~Fechner, and R.~F.~Kiefl,
	Observation of a charge-neutral muon-polaron complex in
        antiferromagnetic Cr$_2$O$_3$, 
	Phys.~Rev.~X {\bf 10}, 011036 (2020). 


\bibitem{yang13} S.~Yang, A.~T.~Brant, N.~C.~Giles, and
	L.~E.~Halliburton,
	Intrinsic small polarons in rutile TiO$_2$, 
	Phys.~Rev.~B {\bf 87}, 125201 (2013).

\bibitem{brant11} A.~T.~Brant, S.~Yang, N.~C.~Giles, and
	L.~E.~Halliburton,
	Hydrogen donors and Ti$^{3+}$ ions in reduced TiO$_2$ crystals, 
	J.~Appl.~Phys. {\bf 110}, 053714 (2011).


\bibitem{shimomura15} K.~Shimomura, R.~Kadono, A.~Koda, K.~Nishiyama,
	and M.~Mihara,
	Electronic structure of Mu-complex donor state in rutile
	TiO$_2$, 
	Phys.~Rev.~B {\bf 92}, 075203 (2015).


	

\bibitem{brown83} J.~A.~Brown, R.~H.~Heffner, M.~Leon, S.~A.~Dodds,
	T.~L.~Estle, and D.~A.~Vanderwater,
	Double electron-muon resonance experiments on muonium in quartz, 
	Phys.~Rev.~B {\bf 27}, 3980 (1983).
\bibitem{estle83} T.~L.~Estle and D.~A.~Vanderwater,
	Theory of double electron-muon resonance, 
	Phys.~Rev.~B {\bf 27}, 3962 (1983).
\bibitem{blazey86} K.~W.~Blazey, T.~L.~Estle, S.~L.~Rudaz, E.~Holzschuh,
	W.~K\"{u}ndig, and B.~D.~Patterson,
	Double electron-muon resonance of anomalous muonium in silicon, 
	Phys.~Rev.~B {\bf 34}, 1422 (1986).
\bibitem{lord04} J.~S.~Lord, S.~F.~J.~Cox, H.~V.~Alberto,
	J.~Piroto~Duarte, and R.~C.~Vil\~ao,
	Double-resonance determination of electron $g$-factors in muonium
	shallow-donor states, 
	J.~Phys.: Condens.~Matter {\bf 16}, S4707 (2004).
\bibitem{doll25} A.~Doll, C.~Wang, T.~Prokscha, J.~Dreiser, and
	Z.~Salman,
	Coherent microwave control of coupled electron-muon centers, 
	Phys.~Rev.~Research {\bf 7}, 033059 (2025).


\bibitem{kroger56} F.~A.~Kr{\"o}ger and H.~J.~Vink,
	Relations between the concentrations of imperfections in
	crystalline solids,
	Solid~State~Phys. {\bf 3}, 307 (1956).
\bibitem{norby10} T.~Norby,
	 A Kr{\"o}ger--Vink compatible notation for defects in
	inherently defective sublattices, 
	J.~Korean~Ceram.~Soc. {\bf 47}, 19 (2010).
	

\bibitem{kojima14} K.~M.~Kojima, T.~Murakami, Y.~Takahashi, H.~Lee,
	S.~Y.~Suzuki, A.~Koda, I.~Yamauchi, M.~Miyazaki, M.~Hiraishi,
	H.~Okabe, S.~Takeshita, R.~Kadono, T.~Ito, W.~Higemoto,
	S.~Kanda, Y.~Fukao, N.~Saito, M.~Saito, M.~Ikedo, T.~Uchida, and
	M.~M.~Tanaka,
	New $\mu$SR spectrometer at J-PARC MUSE based on Kalliope detector, 
	J.~Phys.: Conf. Ser. {\bf 551}, 012063 (2014).

\bibitem{strasser18} P.~Strasser, A.~Koda, K.~M.~Kojima, T.~U.~Ito,
	H.~Fujimori, Y.~Irie, M.~Aoki, Y.~Nakatsugawa, W.~Higemoto,
	M.~Hiraishi, H.~Li, H.~Okabe, S.~Takeshita, K.~Shimomura,
	N.~Kawamura, R.~Kadono, and Y.~Miyake,
	Status of the New Surface Muon Beamline at J-PARC MUSE, 
	JPS~Conf.~Proc. {\bf 21}, 011061 (2018).



\bibitem{kreitzman95}	S.~R.~Kreitzman, B.~Hitti, R.~L.~Lichti,
	T.~L.~Estle, K.~H.~Chow,
	Muon-spin-resonance study of muonium dynamics in Si and its
	relevance to hydrogen, 
	Phys.~Rev.~B {\bf 51}, 13117 (1995).
\bibitem{scheuermann97}	R.~Scheuermann, L.~Schimmele, J.~Schmidl,
	J.~Major, D.~Herlach, and C.~A.~Scott,
        Radio-frequency muon spin resonance (RF$\mu$SR) experiments on
        condensed matter, 
	Appl.~Magn.~Reson. {\bf 13}, 195 (1997).



\bibitem{sm} See Supplemental Material at [URL will be inserted here]
for detailed discussions of the applicability and limitations of the rf 
asymmetry in TF-DEMUR, and the comparison of the $g$ values of Ti$^{3+}$-related centers along $[100]$, $[110]$, and $[001]$.
	
		

\bibitem{brant13} A.~T.~Brant, N.~C.~Giles, and	L.~E.~Halliburton,
        Insertion of lithium ions into TiO$_2$ (rutile) crystals: An
        electron paramagnetic resonance study of the Li-associated
        Ti$^{3+}$ small polaron, 
	J.~Appl.~Phys. {\bf 113}, 053712 (2013).

	
\bibitem{salman14} Z.~Salman, T.~Prokscha, A.~Amato, E.~Morenzoni,
	R.~Scheuermann, K.~Sedlak, and A.~Suter,
	Direct spectroscopic observation of a shallow hydrogenlike donor
	state in insulating SrTiO$_3$, 
	Phys.~Rev.~Lett. {\bf 113}, 156801 (2014).
\bibitem{ito19} T.~U.~Ito, W.~Higemoto, A.~Koda, and K.~Shimomura,
        Polaronic nature of a muonium-related paramagnetic center in SrTiO$_3$, 
	Appl.~Phys.~Lett. {\bf 115}, 192103 (2019). 
\bibitem{ito22} T.~U.~Ito,
        Hydrogen-Ti$^{3+}$ complex as a possible origin of localized 
        electron behavior in hydrogen-irradiated SrTiO$_3$, 
	e-J.~Surf.~Sci.~Nanotech. {\bf 20}, 128 (2022). 


\bibitem{blundell25} S.~J.~Blundell, M.~Bonacci, P.~Bonf\`a,
	R.~De~Renzi, B.~M.~Huddart, T.~Lancaster, L.~M.~Liborio,
	I.~J.~Onuorah, G.~Pizzi, F.~L.~Pratt, and J.~M.~Wilkinson,
	Electronic structure calculations for muon spectroscopy,
	Electron.~Struct. {\bf 7}, 023001 (2025). 
		 
\bibitem{dehn21} M.~H.~Dehn, J.~K.~Shenton, D.~J.~Arseneau,
	W.~A.~MacFarlane, G.~D.~Morris, A.~Maign\'e, N.~A.~Spaldin, and
	R.~F.~Kiefl, Local electronic structure and dynamics of
	muon-polaron complexes in Fe$_2$O$_3$,
	Phys.~Rev.~Lett. {\bf 126}, 037202 (2021).
	
\end{thebibliography}
\end{document}